\documentclass[conference]{IEEEtran}
\newcommand{\chris}[1]{\textcolor{blue}{Chris: #1}}

\newcommand{\fixme}[1]{\textcolor{black}{#1}}

\usepackage{xcolor}
\usepackage[pdftex]{graphicx}
\usepackage[compress]{cite}
\usepackage{hyperref}
\usepackage{ulem}

\ifCLASSINFOpdf
\else
\fi
\usepackage{array}
\usepackage{url}

\newcommand{\university}{University of Illinois at Urbana-Champaign}
\newcommand{\platform}{RokWire} 

\begin{document}
%
\title{Safer Illinois and RokWall: Privacy Preserving University Health Apps for COVID-19 \vspace{-2ex}}




\author{
\IEEEauthorblockN{
\parbox{\linewidth}{\centering
Vikram Sharma Mailthody\IEEEauthorrefmark{1}\IEEEauthorrefmark{3}\IEEEauthorrefmark{2}, 
James Wei\IEEEauthorrefmark{1}\IEEEauthorrefmark{3}\IEEEauthorrefmark{2}, 
Nicholas Chen\IEEEauthorrefmark{1}\IEEEauthorrefmark{3}\IEEEauthorrefmark{2},  
Mohammad Behnia\IEEEauthorrefmark{1}\IEEEauthorrefmark{3}\IEEEauthorrefmark{2}, 
Ruihao Yao\IEEEauthorrefmark{1}\IEEEauthorrefmark{3}\IEEEauthorrefmark{2},
Qihao Wang\IEEEauthorrefmark{1}\IEEEauthorrefmark{3}\IEEEauthorrefmark{2}, 
Vedant Agrawal\IEEEauthorrefmark{1}\IEEEauthorrefmark{3}\IEEEauthorrefmark{2}, 
Churan He\IEEEauthorrefmark{1}\IEEEauthorrefmark{3}\IEEEauthorrefmark{2},
Lijian Wang\IEEEauthorrefmark{1}\IEEEauthorrefmark{3}\IEEEauthorrefmark{2},
Leihao Chen\IEEEauthorrefmark{1}\IEEEauthorrefmark{3}\IEEEauthorrefmark{2},
Amit Agarwal\IEEEauthorrefmark{1}\IEEEauthorrefmark{3}\IEEEauthorrefmark{2},
Edward Richter\IEEEauthorrefmark{1}\IEEEauthorrefmark{3}\IEEEauthorrefmark{2}, 
Wen-Mei Hwu\IEEEauthorrefmark{3}\IEEEauthorrefmark{2},
Christopher W. Fletcher\IEEEauthorrefmark{3},
Jinjun Xiong\IEEEauthorrefmark{2},
Andrew Miller\IEEEauthorrefmark{3}\IEEEauthorrefmark{2},
Sanjay Patel\IEEEauthorrefmark{3}\IEEEauthorrefmark{2}
}
}
\IEEEauthorblockA{\IEEEauthorrefmark{2}C3SR, IBM Research 
\IEEEauthorrefmark{3}University of Illinois at Urbana-Champaign} 
\IEEEauthorblockA{\IEEEauthorrefmark{1}Contributed Equally. Corresponding Email: \{vsm2, jinjunx, soc1024, sjp\}@illinois.edu}
}


%


\IEEEoverridecommandlockouts
\makeatletter\def\@IEEEpubidpullup{6.5\baselineskip}\makeatother
\IEEEpubid{\parbox{\columnwidth}{
    To appear in the Workshop on Secure IT Technologies against COVID-19 (CoronaDef) 2021 \\
    www.ndss-symposium.org
}
\hspace{\columnsep}\makebox[\columnwidth]{}}

\maketitle

\begin{abstract}
COVID-19  has  fundamentally  disrupted  the  way  we  live.
Government bodies, universities, and companies worldwide are rapidly developing technologies to combat the COVID-19 pandemic and safely reopen society. 
Essential analytics tools such as contact tracing,  super-spreader event detection, and exposure mapping require collecting and analyzing sensitive user information.
\fixme{The increasing use of} such powerful data-driven applications necessitates a secure, privacy-preserving infrastructure for computation on personal data.

In this paper, we analyze two such 
computing infrastructures under development at the \university{} to track and mitigate the spread of COVID-19. 
First, we present Safer Illinois, a system for decentralized health analytics supporting two applications 
\fixme{currently deployed with widespread adoption}: digital contact tracing \fixme{and COVID-19 status cards}. 
Second, we introduce the RokWall architecture for privacy-preserving centralized data analytics on sensitive user data. 
We discuss the architecture of these systems, design choices, threat models considered, and the challenges we experienced in developing production-ready systems for sensitive data analysis.
\end{abstract}

\section{Introduction}
COVID-19 has fundamentally disrupted the way we live.
Countless organizations, including government bodies, academic research groups, and companies, are developing and deploying technological solutions to combat the spread of COVID-19~\cite{tcn, dp3t,  google, apple, bluetrace, cmudecentralizedisntriskfree, covidsafepath}.
Worldwide efforts have reinforced that data must play an integral role for safely reopening our communities.  Technologies such as digital contact tracing, superspreader event detection and tracking, exposure mapping, migration mapping, live queues at testing locations, risk assessment, and effective stress management~\cite{digitalcontacttrace} have been developed to help better understand and mitigate the spread of disease.
These techniques require the collection of sensitive user information, introducing a delicately balanced trade-off between data driven functionality and personal privacy.
As more user information is disclosed, the application can provide more accurate, responsive, and personalized experiences; yet the privacy risk increases accordingly~\cite{cmudecentralizedisntriskfree}.
This necessitates trustworthy and secure mechanisms to reduce the risk of compromising sensitive information~\cite{privacyconsideration,canttrustgovtprivacy}.

We believe 
\fixme{that universities} can play a crucial role in this area as they are viewed as relatively trustworthy entities~\cite{CSBS}.
University-led apps can create legitimate trust by establishing public auditors and thorough review processes. 
Furthermore, universities are not reliant on monetizing private data.  
We expect this credibility to encourage widespread adoption.

In early summer 2020, \university{}
announced plans to resume on-campus instruction for the fall semester. 
In order to reach this ambitious goal, the university has taken several initiatives, including the development of technologies for managing the spread of COVID-19 using the \university{} \platform{} platform~\cite{rokwire}. 
Started in 2018, \platform{}'s goal is to serve as an open-source
platform for smart communities, such as campuses, cities, and organizations. 
The prime directive of \platform{} is to provide valuable functionality to users while enabling fine-grain control of their data.
\platform{} does not monetize individual user data and is audited by public authorities.
With the emergence of COVID-19, we 
envisioned that \platform{} should become a platform for a scalable, privacy-preserving computing infrastructure.


In this paper, we detail two secure, privacy-preserving systems developed in the \platform{} platform.  
\fixme{First, we describe Safer Illinois, a system 
for decentralized health analytics and computation, focusing on two of its applications successfully deployed with strong adoption: digital contact tracing based on the recently released Google/Apple protocol and mobile status cards displaying COVID-19 risk.}
We have overcome significant implementation hurdles to develop a scalable solution, addressing significant gaps in existing protocols. 
We provide details on technical challenges, remaining shortcomings, and integration into a broader campus workflow.

Safer Illinois's decentralized architecture enables secure and anonymous digital contact tracing, but also limits analytical potential, particularly in aggregated computation.
Unfortunately, such population-scale insights are critical to forming responsive strategies for pandemic management and public policy.
To address this limitation, we describe RokWall, a generalizable system that can perform centralized privacy-preserving analytics on sensitive user data. 
RokWall enables advanced analytics such as superspreader event detection, exposure mapping, and risk assessment with strong security and privacy guarantees.


We discuss the overall architecture of RokWall, considering both the Intel SGX platform~\cite{intelsgx} and AWS Nitro Enclaves~\cite{nitro}, and detail several different threat models considered.

\fixme{We have successfully deployed Safer Illinois within \platform{} and released it to university members in advance of the Fall'2020 semester at \university{}. We present some early Safer Illinois app usage statistics in this paper. While the data collected so far is limited, we already observe strong adoption and acceptance of Safer Illinois app among the campus community. Even though Safer@Illinois is an entirely optional service, we measure that approximately 82.5\% of the campus population have used the app at least once during Fall'2020. Furthermore, we measure that 53\% of these users also voluntarily opt-in to the contact tracing application.  
Note that this figure likely underestimates the true adoption rate, as the university  students  currently living away from campus should  not  be expected to use the app but are counted in the total university population. }

\fixme{
The RokWall infrastructure is still undergoing active development and not yet available for public use.
During the RokWall architecture implementation process, we encountered and continue to face several technical challenges such as early-stage tool chains, limited availability of trusted execution enviroment (TEEs) in the cloud and the lack of COVID-19 specific datasets. To address this, we are actively collaborating with industry and the Initiative for Cryptocurrencies and Contracts (IC3) to develop tools for TEEs. In particular, we have noticed that there is a lot of pending innovation in enclave tool chains and encourage the community to further explore this segment. }

To summarize, we make the following main contributions:
\begin{enumerate}
    \item Safer Illinois, a \fixme{decentralized} computation system, currently supporting a digital contact tracing application for privacy-preserving exposure notification \fixme{and mobile COVID-19 status cards.} 
    \item RokWall, an 
    architecture for secure, privacy-preserving computing using secure enclaves.
    \item Discuss several technical challenges we face in developing secure, privacy-preserving computing systems. 

\end{enumerate}



We hope this paper fosters discussion 
on developing a privacy-preserving computing infrastructure within the research community.
\section{Safer Illinois: Decentralized Computation}
\label{sec:Safer Illinois}

Exposure notification technologies have become integral components of public health strategies worldwide to curb the spread of COVID-19 infections, often as a digital supplement to manual contract tracing. 
Early successes at staving off the virus in South Korea and Singapore prompted researchers worldwide to develop protocols for effective contact tracing through smartphone devices without significantly compromising individual privacy. 
As with other public health strategies to combat the pandemic, such as facemasks and social distancing, exposure notifications rely on high community adoption rates. 
Simulation-based studies estimate that nearly 60\% of individuals within a region need to be actively using digital exposure notification in order to be effective~\cite{digitalcontacttrace}. 
Our goal with the \platform{} project was to develop an exposure notification solution that could be deployed at scale to around 100,000 users within the University mobile app. The University requested a production ready system by August 2020 to inform public health policies throughout the Fall semester.

Safer Illinois is built around a simple concept: it holds a digital version of your COVID-19 health status.  If you are tested on campus, or by a provider in the surrounding community, the results are stored on your mobile device.   The app then manages the test results by invalidating them after a certain time period determined by county health officials, say 4 days, prompting the user to get re-tested~\cite{uiucsalviatest}.  The results can also be invalidated by a recent encounter with someone whom is later determined to have been infectious at the time, through digital exposure notification.  \fixme{In addition to digital contact tracing, Safer Illinois provides mobile status cards displaying a user's COVID-19 exposure risk.}  To enter a University space, for example, you might be asked to present your digital health status to show that you pose minimal infection risk to others.  Those who opt-out would be asked to show test results by paper or digital image~\cite{saferdoc}\footnote{Please refer \cite{saferdoc} for official University policy.}. 

As security-conscious consumers ourselves, we adopted a privacy-centric philosophy from the onset.  We chose decentralized, privacy-preserving protocols when available. We keep our codebase open-source, for additional transparency~\cite{safer}.  We adopted a minimal data policy, gathering as little data as possible to meet the functionality of the application.

The Safer Illinois architecture involves five components:  (a) exposure notification, (b) integration with testing facilities, (c) administration panel for public health authorities, (d) upload server for positive diagnoses \fixme{and (e) COVID-19 status cards}.   The complexity of our system is primarily in the exposure notification system, so we will focus discussion in this paper on that component, with briefer discussions on the others.

The design space for exposure notification includes a choice of proximity estimation (i.e, Bluetooth, WiFi, ultrasonic, GPS, etc), centralized vs. decentralized vs. hybrid architecture, cryptographic
protocol, etc.  Our approach was to leverage the ongoing work by various security experts and communities worldwide, who were creating open-source protocols for digital exposure notification. 

We evaluated three protocols in depth, namely the Temporary Contact Number (or \texttt{TCN}) protocol~\cite{tcn}, Decentralized Privacy-Preserving Proximity Tracing (or DP-3T)~\cite{dp3t} and the Google/Apple Exposure Notification (or GAEN)  protocol~\cite{google,apple}, each of which were summarized briefly below.

\textbf{TCN protocol} generates a Temporary Contact Number (or \texttt{TCN}), a psuedo-random identifier derived from a seed, every 15 minutes.  Unique \texttt{TCNs} are exchanged via Bluetooth Low Energy (or BLE) and stored when two devices come in close proximity. 
When a user tests positive, a report is sent to a centralized server with the list of \texttt{TCN}s exposed.  User devices pull this report and determine matching \texttt{TCN} to see if the user has been exposed. 

\textbf{DP-3T protocol} 
differs from the TCN protocol on how anonymous IDs are generated (random seed with deterministic hash + truncation vs asymmetric key-pair with deterministic hash ratchet in TCN), what information is reported (\texttt{EphID} and seed of all relevant epochs vs public key with start and end time and \texttt{tck} for regenerating \texttt{TCN} for timeblock) and what information is stored (\texttt{hash(EphIDs)} and epoch \texttt{i}, proximity, duration and coarse time indication vs \texttt{TCN} value).   

\textbf{GAEN protocol} shares concepts from the DP-3T and TCN protocols, including the use of BLE for proximity detection, with key differences in anonymous ID generation (Rolling Proximity Identifiers (\texttt{RPIs}) generated through Temporary Exposure Keys (\texttt{TEKs}) every 10 minutes) 
and reporting of positive test cases (\texttt{TEK}s and a timestamp represented as an epoch interval number). Unlike DP-3T and TCN, the GAEN protocol is publicly described, but is still partially closed-source. Access to the implementations are only granted to public health authorities operating at the state or country-level; at the time of this writing, they were not available to our team. 


We evaluated these protocols in April and May 2020, a time when these concepts were still undergoing intense development and existing codebases were not yet mature.  The open-source code had known shortcomings, such as failing in BLE background mode for iOS devices.  We decided to adopt the GAEN approach and build our own implementation, \fixme{ while planning to switch to the Google/Apple implementation} in case we received API entitlements from Google and Apple due to our affiliation with a large University\footnote{We have not yet received such entitlements}.  

In the overall user workflow of the app, an individual can get tested on campus using one of several testing sites. 
As a sidenote, we employ a breakthrough saliva-based test developed at \university{} that enables high-throughput testing of up to 10,000 tests per day at low cost~\cite{uiucsalviatest}.
The user presents their University ID when a test is administered, thereby linking their results to a University ID number.  Since the user must authenticate within the app using their University credentials,  their test results can be linked to the user via the app.  The user is notified by the app once the test results are available, typically within an day. Test results can be encrypted using the user's public key and pushed onto the user's device with the user's consent. 

If the diagnosis is positive, the user can choose to upload a history of their \texttt{TEKs} to a diagnosis upload server. Apps with exposure notification enabled will periodically download published TEKs from the diagnosis upload server, decode the TEKs into rolling proximity identifiers, and check for matches with \texttt{RPIs} stored in the local device database.  As a further security measure, the upload server will use one-time codes that are electronically shared with the testing sites.  A single code is provided alongside each test result to the user device, which is then used to establish 
a chain-of-authenticity from the testing
\fixme{site} to the upload server, via the user device.

If a matching \texttt{RPI} is found, an exposure score is calculated using parameters such as duration of exposure, reception and transmission strength of the Bluetooth signal, an estimated onset date of infection, and models of testing efficacy.  How such parameters can be used to estimate the risk of infection is an ongoing area of work both within \university{} and elsewhere~\cite{covidsafepath}. \platform{} contains an admin control panel 
\fixme{that provides} public health authorities \fixme{ with} 
limited ability to adjust the parameter weighting system used to score an exposure.  If the score is above the threshold, indicating exposure risk, then the user's most recent test result is invalidated, prompting the user to be retested and setting their mobile status card to reflect high risk.

Complementing this workflow is the exposure notification functionality, running continuously on each device.  Safer Illinois directly follows the specification defined by the GAEN protocol in generating and exchanging exposure keys. Every day, each user generates a unique Temporary Exposure Key which constructs a user's Rolling Proximity Identifier Key and subsequent \texttt{RPIs} to be exchanged with other users.  In addition, the \texttt{TEK} generates an Associated Encrypted Metadata (\texttt{AEM}) Key which, along with an \texttt{RPI}, can be used to encrypt a few bytes worth of optional metadata. 

Each user broadcasts their \texttt{RPI} and corresponding \texttt{AEM} with a rolling period of approximately 10 minutes. Whenever a contact is registered within the range of the device's effective Bluetooth range, the device saves the detected \texttt{RPI}, contact duration and Bluetooth received signal strength - known as RSSI - to local storage. The device also securely saves the user's daily \texttt{TEK} and a timestamp to be uploaded to a server in case the user tests positive for COVID-19.

\subsection{Security}
The security and privacy implications of 
exposure notification protocols have been heavily examined by experts, including the DP3T and TCN communities~\cite{dp3t,tcn}.  We briefly summarize the salient threat models \fixme{ that represent potential vectors for attackers to learn the identity of other users involved in a contact exchange or positive test result.}
We separate these threats into two categories: 1) \textit{inherent attacks} faced by all Bluetooth proximity tracing systems, and 2) \textit{protocol-dependent attacks} which depend on how the protocol generates and exchanges its anonymous identifiers.

\textit{Inherent security considerations:} When a user is notified of an exposure event, they may be able to identify the infected individual by correlating their interactions with the reported time of exposure.
Even if the application obfuscates the timing with noise, an attacker can create multiple accounts or use multiple phones at different times \fixme{ to cancel or reduce the noise introduced by the system}.
This threat compounds further if attackers log additional interaction data from infected persons or triangulate data from third-party sources, such as building access logs. 
%
Moreover, apps that solely rely on Bluetooth to exchange keys can be susceptible to certain broadcasting threats. 
If an attacker were to set up powerful transmitters to enhance their effective Bluetooth range, false contacts could be logged. 
Alternatively, an attacker may set up a Bluetooth jammer that could disrupt communication between devices.

\textit{Protocol-dependent security considerations:}  
To begin with, anonymous identifiers must not be linkable to one another nor to the transmitting device.
The former is achieved in all protocols discussed through cryptographic pseudorandomness while the latter requires the synchronization of rotations of Bluetooth MAC address and anonymous identifier. 
Additionally, there remains a threat of replay attacks, where adversaries record anonymous identifiers in one area and replay them in another location causing public disruption or targeting specific individual or area. 
A solution to this problem is to allow the attacker to duplicate and transmit identifiers, but inhibit notification to users who receive these fraudulent signals \cite{contrail}. 
All three protocols mitigate this issue to some extent by incorporating timestamps while checking for exposed matches. 






\subsection{Implementation Challenges}

We designed our approach to exposure notification with an emphasis on wide-scale deployment.  
Ideally, the protocol could be adopted with minimal impact to users by providing ease of use, minimal energy consumption and privacy and security guarantees.
We picked the Google/Apple API for a variety of reasons.  The API was more stable at the time we were examining the various alternatives, and early experience with the protocol would pave the path if we were later granted entitlements to use the API within our application.  When we embarked on the project, we did not have entitlements to the GAEN API, so we set out to develop our own implementation of the protocol while addressing the known issues suffered by DP3-T and others.
%
%
Below, we describe some of the challenges encountered in implementing a scalable, production-ready system at the application level. 

\textbf{iOS Background Advertising:}
Moving an iOS application to background mode restricts its Bluetooth advertisement packets.
Namely, instead of advertising a standard service UUID, transmissions are moved to an ``overflow area" where they are only observable by a device explicitly scanning for it. 
Since all iOS background apps \fixme{ on the same device} share the same overflow area, there is no guarantee that the app is advertising a preset bitmask. 
Moreover, there is a possibility of collision if two Bluetooth services from different apps set the same bitmask, thus an app may detect a different service than intended. 
Currently, we do not have a solution to this problem; however, the likelihood of 
\fixme{ such conflict} is very low, as few other apps (if any) advertise Bluetooth in the background. 

\textbf{iOS-iOS Background Communication:}
In Android, a callback can be set up to detect the overflow bitmask of an iOS background device. 
On iOS devices, however, this callback would only be triggered if the screen is turned on and beacon ranging is enabled. We found this can be circumvented by sending a local notification, which will illuminate the screen for 10 seconds at the expense of battery life. 

\textbf{Bluetooth Mac Address Changes:}
It is essential to align Bluetooth MAC rotations with each \texttt{RPI} change. 
Otherwise, an attacker can correlate \texttt{RPIs} coming from a single user. 
Unfortunately, as of Android 6.0 and iOS 8, an application cannot control the timing of its Bluetooth MAC address changes or even identify when this change occurs. 
\fixme{However, we found that the Bluetooth MAC address changes every time the advertising service restarts on Android. 
We took advantage of this finding in our Android implementation by restarting the advertising service to obtain a new MAC address every time a new \texttt{RPI} is generated.
Unfortunately, we did not observe a similar phenomena for iOS, and it remains an unsolved issue.}

\textbf{iOS Background Execution:} 
With iOS devices, we found it difficult to keep an app from being suspended by the OS when in background mode.  Suspended apps will not be able to record or transmit \texttt{RPIs}. 

\textbf{Battery Efficiency:}
%
Constant Bluetooth scanning and advertising takes a substantial toll on battery life. While the GAEN protocol sets scanning intervals at 5 minutes apart, Android and iOS SDKs provide little control over these intervals. Android provides 3 scan settings, but the actual times may differ by manufacturer, while no such options are documented for iOS. Even in the most conservative power-saving state, scanning occurs every couple seconds. Table~\ref{tab:battery} shows the average battery drain in \% per hour across iOS and Android. We set Safer Illinois to be the only application running with it constantly scanning another device. However, these numbers may vary by other factors including device usage, other devices scanned, OS level, and device model. 
\fixme{From Table~\ref{tab:battery}, the Safer Illinois app consumes reasonable amount of energy over time depending on the OS and device. }

\begin{table}[]
    \centering
    \caption{Average Battery Drain per hour of Safer Illinois}
    \label{tab:battery}
    \resizebox{0.8\linewidth}{!}{%
    \begin{tabular}{c|c|c} 
    \textbf{Device} & \textbf{Safer Illinois On}  & \textbf{Safer Illinois Off} \\
    Google Pixel 3    &    1.12\%           &       0.47\%    \\
    iPhone X &    4.8\%        &      0.59\%     \\

    \end{tabular}
    }
    \vspace{-4ex}
\end{table}






\section{RokWall: Centralized Enclave Computation}
\label{sec:rokwall}

\fixme{The Safer Illinois application demonstrates privacy-preserving computation on sensitive user data within a decentralized framework.}
However, desirable services such as exposure mapping, secure data transfer, and safety status verification require centralized analysis.
In comparison with decentralized implementations, a centralized infrastructure requires users to place greater trust in service providers' benevolence and honesty.
While reasonable for a highly transparent university organization, users may justifiably remain skeptical of private businesses or other third parties accessing their data within the \platform{} system.
Thus, we required a secure, privacy-preserving computing infrastructure inside \platform{} for centralized analytics. 

To satisfy this need, RokWall is guided by the following fundamental principles:
\textit{(a) Privacy:} 
Sensitive user data is only used by services authorized by the user. Users have assurance that a third-party service provider cannot exploit beyond the declared capabilities. 
\textit{(b) Security:} 
No party, including service providers and manufacturers, can access data beyond the computation's output, and
\textit{(c) Accountability:}
Users or public auditors can review the code bases, verify program binaries and ensure it meets all security and privacy guidelines. 

\begin{figure}[h]
    \centering
    \includegraphics[width=\linewidth]{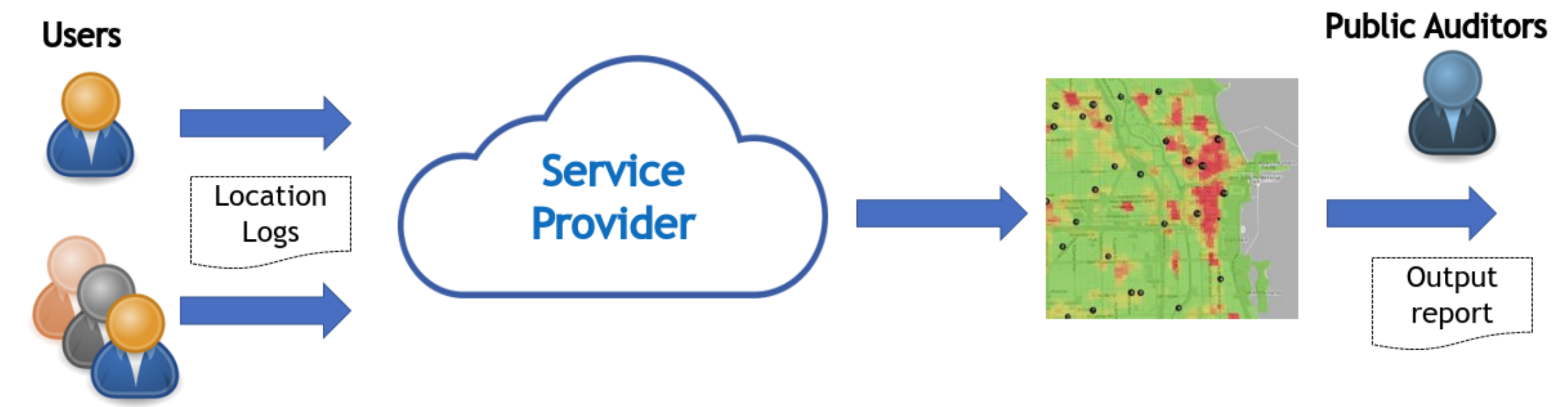}
    \vspace{-4ex}
    \caption{Exposure mapping application on sensitive location data. User uploads sensitive location logs to the service provider. The service provider generates an exposure map along with an output report for public auditors to audit.}
    \label{fig:exposuremap}
\end{figure}

\textit{Exposure Mapping Application:}
We present a COVID-19 exposure mapping application in Figure~\ref{fig:exposuremap} as an example of privacy-preserving computation on sensitive data.
Exposure mapping aggregates user location data to calculate a heat map, visualizing the risk of infection exposure. 
This application helps health authorities assess the likelihood of superspreader events and warn the general public of high risk areas.
GPS location data is highly sensitive, so the service provider should follow previously discussed fundamental principles: \textit{(a)} Perform only a minimal set of queries on the user's location data to preserve privacy,
\textit{(b)}  Ensure the data is secure and used only for exposure mapping application purposes, and 
\textit{(c)} Enable auditors to verify these guidelines with public information such as output report to hold the service provider accountable.

To this end, we present RokWall, a secure architecture (see $\S$~\ref{sec:rokwallarch}) for sensitive data computation. 
We apply RokWall to COVID-19 exposure mapping while preserving the desired security and privacy guarantees for user location information.
We analyze various threat models  (see $\S$~\ref{sec:rokwallthreatmodel}) considered for the exposure mapping application and RokWall's protection against various attack vectors.  
Finally, we present various technological challenges (see $\S$~\ref{sec:rokwalltc}) faced during deployment and provide  potential solutions.

\subsection{RokWall Architecture}
\label{sec:rokwallarch}
\begin{figure}[t]
    \centering
    \includegraphics[width=\linewidth]{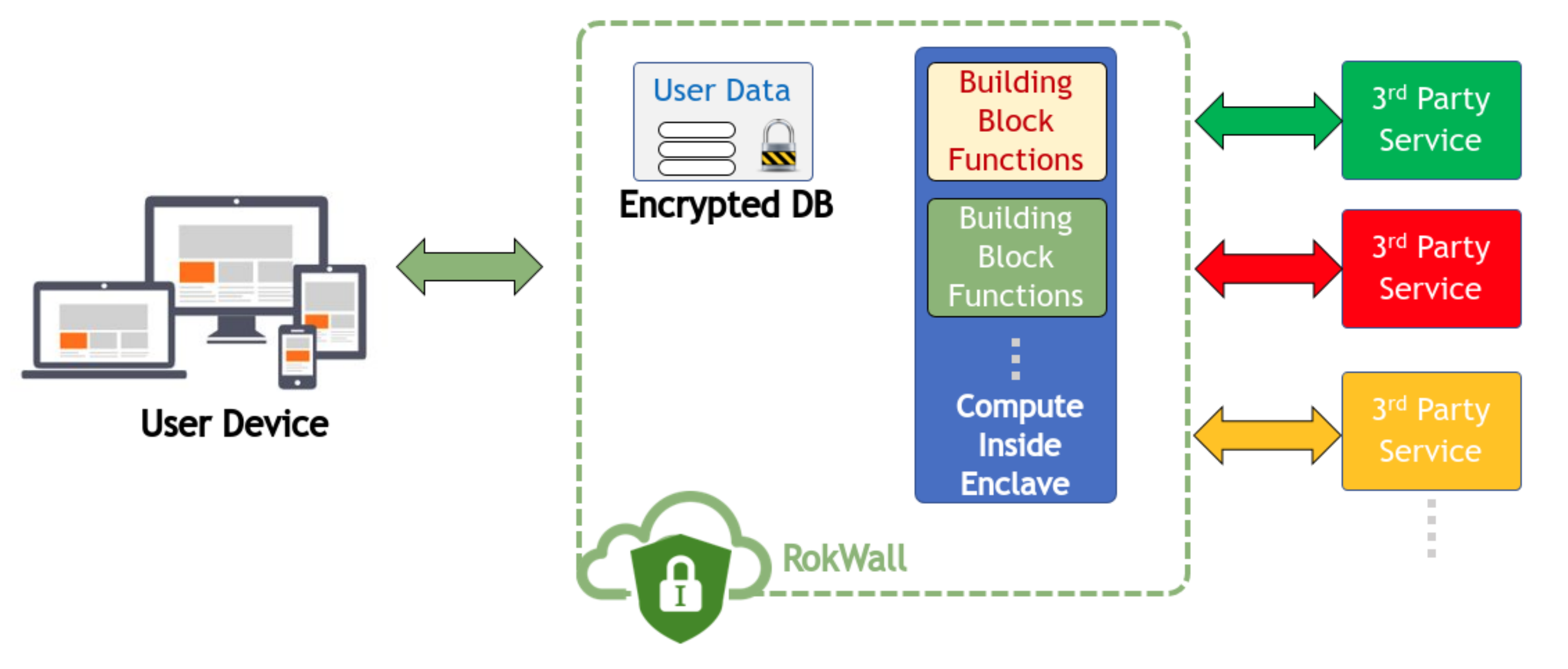}
    \vspace{-5ex}
    \caption{Overall strawman architecture of RokWall is shown.}
    \label{fig:rokwallarch}
    \vspace{-3ex}
\end{figure}

\fixme{Guided by the preceding principles of security, privacy, and accountability, we primarily considered two established solutions for secure computation: multi-party computation (MPC) and enclave based trusted execution environments (hardware~\cite{intelsgx}/software~\cite{nitro} enclaves)
While both options could potentially satisfy our required standards, we noted that enclave tool chains were considerably more mature and production-ready~\cite{songsgxmpc,conclave}.
Due to the time sensitive nature of our mission and performance requirement, we decided to proceed with an enclave-based architecture. In the future, we may reevaluate the merits of MPC and consider supporting it as an alternative.}




Figure~\ref{fig:rokwallarch} provides a high level illustration of the RokWall architecture using secure enclaves.
The RokWall architecture supports sensitive data computation by leveraging secure enclaves.
\fixme{We considered two trusted execution environments (TEEs) - Intel's SGX, a hardware enclave, and the more recent AWS Nitro software enclave~\cite{nitro}. 
In theory, the choice of platform may be application specific, 
as each platform imposes different security/performance tradeoffs.
RokWall will eventually support both of these platforms, enabling use case specific choice. 
However, we will limit the threat model discussion to the Intel-SGX platform as it is currently more established and thoroughly vetted. }



All data analysis, such as exposure mapping or COVID-19 risk calculation, occurs entirely within the confines of the secure enclaves. 
Each individual data analytics function is referred to as a ``building block function" and are statically linked with enclaves.
Each building block function publicly declares a hash of its program binary and each secure enclave generates an output public key. 
Source code of all the building block functions and APIs are planned to be open sourced and thoroughly audited. 

RokWall allows third-party services to upload information, such as a health authority's API updating test results of a specific user using a secure channel. 
Users upload sensitive data to the RokWall server using a secure encrypted channel such as Transport Layer Security (TLS) along with the enclave public key and hash of program binary for the user authorized application. 
%
Inside RokWall, user data is stored in an encrypted database. 
During the query execution, only the building block function or third-party services whose hash of the program binary matches with the user-approved application can temporarily decrypt and access data within the secure enclave.
Critically, this guarantees that unencrypted data never leaves an enclave.
An unauthorized building block execution will result in the generation of a useless result. 


\textit{Remote attestation in RokWall:}
Remote Attestation allows cryptographic verification of the code allegedly executed inside an TEE. 
RokWall uses a 3-party EPID based remote attestation mechanism for Intel-SGX hardware~\cite{intelsgx} \fixme{and the AWS KMS service for Nitro enclaves~\cite{nitro}. }
We reduce verification effort needed for end-user devices by publicly providing verified attestation report generated by the RokWall enclave. 
This report would contain information about the enclave code (given by MRENCLAVE in case of SGX) as well as the public-private key pair generated during enclave initialization. Auditors (or even users) can verify that the MRENCLAVE information in the report matches the publicly available MRENCLAVE generated by building/compiling the enclave code, vetted by interested parties. \fixme{The procedure is similar when RokWall uses Nitro enclaves instead of SGX.}

\textit{Exposure Mapping Function in RokWall:}
RokWall uses the Intel SGX platform to provide a trusted execution environment for the exposure mapping application.  
Users upload sensitive location logs using TLS to the RokWall encrypted database along with a hash of the exposure mapping binary and the enclave's public key. 
On a regular interval (in this example, once per day), the exposure mapping enclave generates and publishes a heat map as output on a public server along with a signature and report for remote attestation. 
The user app can pull this output from the public server, then verify the signature. 


\subsection{End-to-end Chain Of Trust In RokWall} 
\label{sec:rokwallthreatmodel}

Security and privacy guarantees are primary principles of the RokWall design. 
We consider a three-tiered threat model: 
(1) \textit{network attackers}, (2) \textit{client attackers} and (3) \textit{service provider attackers}. 
To safeguard against network attackers, clients communicate with the RokWall server via TLS channel. 

Unfortunately, we cannot currently prevent client attackers from running malicious code or flooding the system with spoofed data. 
This is a known problem on systems that do not require user verification.
One possible solution, employed by electronic voting systems~\cite{evoting}, allows an authority to register public keys of users.
University officials could distribute public keys to community members interested in using the service.

Service provider attackers can be classified into three sub-categories: \textit{(1) server-software}, where a service provider runs malicious user-level software, \textit{(2)  server-kernel}, where a service provider runs malicious kernel-level software, and \textit{(3) server-hardware}, where a service provider has physical access to the server hardware. 

\textbf{Server-software attacks}: 
Server-software level attacks assume that the service provider is limited to user-level privileges.
This includes writing and running malicious code, but excludes kernel privileges or hardware attacks.
Server-software attacks can generally be prevented by using SGX enclaves. 
Remote attestation enforces transparency and enables public auditors to review code, while data sealing ensures that the service provider cannot access raw, decrypted user data.
One remaining attack is an isolation attack, where a service provider runs the query with only a single victim user's location logs.
This query yields a heatmap exposing the victim's location history, even though the code would pass an audit.

RokWall addresses this with a two step solution: 
\textit{(1)} employ non-volatile counters such that a location log can only be used for a heatmap one time, and
\textit{(2)} output a hash of location logs included so a user can verify that their data was used for generating a given heatmap.
Then, if a service provider commits an isolation attack, the victim's data will necessarily not be present in the official heatmap. 
If a user finds that their data is not present in an officially published heatmap, they can then report the service provider to RokWall administrators. 

\textbf{Server-kernel attacks}: 
Server-kernel attacks expand upon user-level code execution and permit the attacker to inspect memory management within SGX.  This level of attack can theoretically allow privileged side channel attacks, exposing memory access patterns even in sealed data~\cite{serversidechannel}. 
We avoid leaking information to these attackers by ensuring data oblivious execution and guaranteeing a constant runtime regardless of input size.
In the case of exposure mapping, this entails unsealing and resealing the entire heatmap every time data is updated.  
RokWall currently does not defend against microarchitectural attacks (like cache-timing attacks) as they pose far more sophisticated adversaries.

\textbf{Server-hardware attacks}: 
Server-hardware attacks involve physically probing or tampering with the enclave's system hardware.
We generally expect the cloud service provider to ensure the physical security of their servers. 
We are still investigating additional counter measures to address these attacks and will address them in the future.

\subsection{Technical challenges}
\label{sec:rokwalltc}
Computation on sensitive data raises a number of practical constraints that manifest when implementing a production-ready system.
We describe some of the challenges we encountered while developing RokWall system and propose solutions.

\textbf{Monotonic counter on Intel Servers:} 
Rollback attacks present a general security problem for enclave solutions.
An adversary OS can restart the service with an outdated version of sealed data and leverage it to leak user information.
Intel provides a native SGX monotonic counter service to tackle this problem, while AWS Nitro enclaves do not support non-volatile counters. However, SGX cloud services such as IBM Cloud and Microsoft Azure are currently built on Intel Xeon E3 server-grade processors, which do not support the Intel Management Engine required for enabling SGX monotonic counter service. Alternatives to SGX's native monotonic counter have been proposed, including distributed rollback protection systems such as ROTE~\cite{rote}.
Other solutions include the migration of the counter service to a third-party trusted source or a BFT distributed network such as CCF~\cite{ccf}. 
\fixme{RokWall uses the CCF network to provide non-volatile monotonic counter support for its enclaves. }

\textbf{SGX Memory management:}
Intel SGX provides data sealing for encrypting and saving confidential enclave information to persistent storage. Sealing comes in two forms, Enclave Identity \fixme{ based} vs. 
\fixme{ Signing} Identity \fixme{ based}. 
Data sealed with Enclave Identity (MRENCLAVE) will only allow other instances of the same enclave to unseal, whereas Signing Identity allow other versions and builds of the enclave to unseal. \fixme{RokWall currently uses Enclave Identity for sealing to prevent successive data encroachment; user authorization should apply to an application as it is currently described.} \fixme{Signing Identity would allow future versions of the enclave signed by the same Signing Identity to access sealed data.} However, Intel SGX sealing is not intended for large data objects. In addition to performance degradation, crossing EPC memory bounds requires memory management from the enclave itself. 

\textbf{Challenges with Remote Attestation:} 
A major challenge in implementing remote attestation is ensuring reproducible builds between auditors, clients and the RokWall server, as inconsistent builds can raise false MRENCLAVE mismatches. 
Furthermore, the auditors (or users) must use identical backend libraries/packages as described in the attestation report in their build process.
In practice, this may pose a significantly inconvenient task for auditors. 
\fixme{Additionally, available tool chains such as containers for remote attestation and reproducible builds are far from production quality, especially for use with the Intel SGX platform. 
To address this unmet need, we are working closely with the Initiative for Cryptocurrencies and Contracts (IC3) to enable reproducible enclave builds for the purpose of TEEs~\cite{teeons}. 
}



\textbf{Testing Dataset:}
When developing the exposure mapping building block, we struggled to find an appropriate, publicly available GPS dataset for simulating infection dynamics. 
We ultimately decided to test RokWall’s location related queries on the T-Drive GPS trajectories data~\cite{tdrive}.
T-Drive records coordinates for 10,000 taxi cabs in Beijing over the course of a week.
Some comparative advantages of the T-Drive dataset are its high number of entities, dense population concentration, and high frequency of reporting.  

While the T-Drive dataset is sufficient for initial testing, it has several key limitations. 
Critically, the data isn’t perfectly representative of our eventual use cases since the entities are vehicles, rather than people. 
Taxis are confined to roads and don’t enter buildings so we cannot run indoor, intra-building analysis. 
Moreover, this data can’t facilitate algorithmic parameter tuning, such as heatmap granularity or super-spreader event thresholds, because of differences in population density and entity size. 
Thus, we will likely need to collect organic human data for fine tuning.

\subsection{Additional RokWall Services}
\label{sec:virtualstatuscard}
\fixme{
In addition to enabling the exposure mapping use case, we are particularly excited by the RokWall infrastructure's potential for wider generalization.  We are currently investigating two additional timely use cases which will depend on RokWall for secure, privacy-preserving computation.}

\fixme{\textbf{Secure Data Transfer:}
During the Safer Illinois deployment process, we encountered a pressing need for a secure data transfer mechanism.
We found that, in practice, users often changed their mobile devices due to upgrades and repairs.
In these situations, a user would need to recover credentials and transfer sensitive personal data, like stored contact tracing RPIs and past test results, between devices.  
To facilitate this process, we currently provide a mechanism to transfer data between two devices using QR codes and Bluetooth. 
However, this design has limitations as it requires both devices to be accessible and functional, which may not necessarily be the case. 
To address this problem, enclave data sealing and remote attestation can enable certifiably secure data storage and retrieval. This use case will require additional investigation, but we hope to enable it in the near future. }

\fixme{\textbf{Virtual Status Card:}
As previously described, the Safer Illinois app is intended to complement frequent testing in minimizing the spread of COVID-19. 
However, it is important to acknowledge that a university community is exceptionally conducive for these technologies due to high tech literacy and device ownership of its inhabitants; unfortunately, expanding operation to the outside world entails a very different set of practical assumptions. 
Notably, consider scenarios in which residents may not have access to a personal mobile device. For example, young children or low income households may not own a mobile phone.
Subject to these limitations, we are exploring methods to build a virtual status card application to determine COVID safety status using untrusted client devices. 
}

\fixme{We believe that RokWall can enable this service in a secure manner. As in the exposure mapping use case, data sealing ensures that plaintext medical records can never be accessed outside of the secure enclave, even by a compromised service provider. 
Furthermore, remote attestation can facilitate credible rate limits or user alert policies.}

\section{Discussion}
\label{sec:open}

\subsection{Safer Illinois Usage Statistics}
\fixme{Following a four month development process, Safer Illinois was deployed at the start of the Fall 2020 semester.  
Here, we present real-world data collected from live community usage.}

\fixme{The \university{} has a campus population of approximately 60,000 \cite{uiucpop}.
Though Safer Illinois is an entirely optional service, we measure that approximately 82.5\% of the campus population have used the app at least once during Fall'2020.  Furthermore, we expect that this number may still be an underestimate of relevant app adoption.
After all, university students currently living away from campus, perhaps due to safety concerns and the current prevalence of online coursework, should not be expected to use the app but are counted in the total university population. 
}

\fixme{
Over the time period of 11/30/2020 to 12/15/20, we recorded that 53\% of these Safer Illinois users had voluntarily enabled exposure notification.
Additionally, this figure is a strict underestimate of the true value because usage data is sampled only when a user undergoes their routine COVID-19 test. Since Safer Illinois consumes a substantial amount of battery, not all users enable exposure notification functionality all the time.  Therefore, users who temporarily disabled exposure notification at time of test are falsely counted as permanently disabling the service.
Please note that the Safer Illinois app and exposure notification enrollment are optional, opt-in services and not required to access any university services.
We are particularly heartened to see such a large portion of the population opt-in voluntarily, demonstrating significant trust and appreciation for efforts responding to COVID-19.
}

\fixme{
During the same time period, we observed 19,439 average unique users using the Safer Illinois app per 4-day interval, either to check the test results or access building services using their status card. 
Because the university requires each member to test once every 4 days (4 days without a test automatically results in status change~\cite{saferdoc}), we present our data as averaged over 4-day intervals to approximate a cross section of the population.
Although the average unique users may initially seem to be a small fraction of the campus population, note that for this period, all classes were held entirely online and many students had left campus.}

\fixme{\textit{In summary, this data demonstrates substantial public interest and acceptance for the digital contact tracing and building access status card services. }
However, we still have to collect more data to determine the efficacy of digital contact tracing and we hope to address this question in the near future. 
}

\subsection{Availability of secure enclaves in the cloud}
\fixme{
Confidential computing infrastructure has been evolving for more than two decades.
For much of that time, though, its general availability and support system with cloud vendors, such as Amazon AWS, Microsoft Azure, Google Cloud and IBM Cloud, was limited.
However, perhaps instigated by the pandemic, we have recently observed burgeoning deployment of secure, privacy-preserving cloud computing services.
Although current tool chains and software are immature, this trend demonstrates marked demand for such solutions.}

\fixme{At the time of writing, IBM Cloud and Microsoft Azure support Intel-SGX~\cite{intelsgx} based hardware TEEs while Google uses the AMD SEV~\cite{amdsev} hardware TEE for confidential computing projects~\cite{msftcc,ibmcc,gcpcc}. 
Meanwhile, AWS software-based Nitro enclaves are both serverless and scalable~\cite{nitro}.
Nitro enclaves provide hardened and constrained virtual machines (VMs). The restricted enclave VM solely interacts with its host instance via a secure local channel. 
Like many hardware enclaves, Nitro provides a cryptographic remote attestation service.
}


\fixme{While the hardware TEEs, such as Intel SGX or AMD SEV, charge an additional price ranging between \$4-\$30 per instance per month, software enclaves typically come at no cost to the developer. Additionally, in contrast with hardware enclaves, Nitro offers flexible computing resource allocation, including memory and CPU cores.}
\fixme{The downside of software enclaves, however, is that they assume a weaker threat model. When using a software enclave, one needs to implicitly trust the service provider for all of remote attestation, data sealing, key management and software infrastructure.}

\subsection{Why design both decentralized and centralized systems?}
Contact tracing can be implemented in a decentralized or centralized fashion, which has traditionally forced developers into making a trade-off between user privacy and analytic capability.
While previous works frequently favor decentralized implementations due to privacy concerns, RokWall can enable centralized \fixme{data analysis} while upholding user privacy.

In order to develop a functional and reliable contact tracing system by start of the Fall 2020 semester, Safer Illinois leveraged preexisting GAEN APIs in a decentralized system.
However, a decentralized architecture carries inherent limitations that can be solved by centralized analysis on user data.
For digital contact tracing, centralized GPS data analysis can help identify infection hotspots, remedy bluetooth connectivity issues, and enable cross-time analysis.
Moreover, centralized systems can absolve reliance on user-owned client devices, as described in Virtual Status Card.
We envision eventually migrating \fixme{parts} of Safer Illinois to RokWall, enabling richer analysis \fixme{and broader functionality. }

\section{Conclusion}
\label{sec:conclusion}

In this work, we introduced Safer Illinois and the RokWall architecture under development in the \university{}'s \platform{} platform. \fixme{Safer Illinois enables privacy-preserving digital contact tracing and COVID-19 status cards with decentralized computation.}
Meanwhile, RokWall presents a general framework upon enclave TEEs for secure, privacy-preserving centralized analytics. We detailed our design choices and threat models considered while implementing a production-ready system. 
We also presented several technological challenges and lessons learned from deploying these systems in practice.
We hope this work fosters discussion in developing a privacy-preserving computing infrastructure.

\section*{acknowledgement}
\fixme{
We would like to thank all of the volunteers and staff of \university{} who helped in many ways during this pandemic. We particularly appreciate the help of the Engineering IT team, the Inabyte team, the NCSA team, Professor John Paul, Professor William Sullivan, Todd Nelson, Nickolas Vance, Isaac Galvan, Edward Delaporte, Tracy Smith, Mary Stevens, Melvin Fenner, Kathryn Courtney, Nerla Jean-Louis and Sylvain Bellemare of IC3. 
This work is partially supported by IBM-ILLINOIS Center for Cognitive Computing Systems Research (C3SR) and University of Illinois. We would also like to thank the workshop organizers, PC chairs and reviewers for making this workshop happen. }

\bibliographystyle{IEEEtranS}
\bibliography{ref}

\end{document}